\documentclass[preprint2]{aastex}

\def\ee{$e^\pm$}
\def\g{$\gamma$}
\def\ginga{{\it Ginga}}

\def\exosat{{\it EXOSAT}}

\begin{document}

\title{Observations of Seyferts by OSSE and parameters of their X-ray/gamma-ray
sources}

\author{Andrzej A. Zdziarski}
\affil{N. Copernicus Astronomical Center, Bartycka 18, 00-716 Warsaw, Poland}
\email{aaz@camk.edu.pl}
\author{Juri Poutanen}
\affil{Stockholm  Observatory, SE-133 36 Saltsj\"obaden, Sweden}
\email{juri@astro.su.se}
\and
\author{W. Neil Johnson}
\affil{E. O. Hulburt Center for Space Research,
Naval Research Laboratory, Washington, DC 20375, USA}
\email{johnson@osse.nrl.navy.mil}

\begin{abstract} We present a summary of spectra of Seyfert galaxies observed
by the OSSE detector aboard {\it Compton Gamma Ray Observatory}. We obtain
average spectra of Seyferts of type 1 and 2, and find they are well fitted by
thermal Comptonization. We present detailed parameter ranges for the plasma
temperature and the Compton parameter in the case of spherical and slab
geometries. We find both the average and individual OSSE spectra of Seyfert 2s
are significantly harder than those of Seyfert 1s, which difference can be due
to anisotropy of Compton reflection and/or Thomson-thick absorption.
\end{abstract}


\section{Introduction}

One of major unresolved problems of astrophysics of active galactic nuclei
(AGNs) is the form of their soft \g-ray spectra (at energies $\ga 100$ keV).
The X-ray spectra of Seyfert galaxies are known relatively well, and consist
(at $\ga 2$ keV) of power-law, Compton reflection, and Fe K$\alpha$ components
(e.g., Nandra \& Pounds 1994; Nandra et al.\ 1997; Zdziarski, Lubi\'nski \&
Smith 1999). However, the form of a high-energy cutoff of the power law remains
poorly constrained. In the case of the brightest radio-quiet Seyfert, NGC 4151,
results from the Oriented Scintillation Spectroscopy Experiment (OSSE; Johnson
et al.\ 1993) aboard {\it Compton Gamma Ray Observatory\/} show the spectrum
above 50 keV is well described by thermal Comptonization (e.g., Zdziarski,
Johnson \& Magdziarz 1996; Johnson et al.\ 1997a, hereafter J97). However, the
X-ray spectrum of NGC 4151 is heavily absorbed and it is not clear whether its
intrinsic spectrum is typical for Seyferts.

In the case of other radio-quiet Seyferts, constraints (from either OSSE, {\it
RXTE\/} or {\it BeppoSAX}) on the form of their individual soft \g-ray spectra
are rather poor due to limited photon statistics. One way to better constrain
the soft \g-ray properties is to consider average spectra. Studies of the
average $\sim 2$--500 keV spectra of Seyferts observed by both \ginga\/ and
OSSE (Zdziarski et al.\ 1995; Gondek et al.\ 1996, hereafter G96) and \exosat\/
and OSSE (G96) have shown that purely non-thermal models (e.g., Lightman \&
Zdziarski 1987) are highly unlikely. Although those authors have shown that
thermal Comptonization provides a good description of the observed spectra,
they have approximated model spectra from that process by a power law with an
exponential cutoff. This simple model has also been used in numerous studies of
individual objects (e.g., Zdziarski et al.\ 1994; Bassani et al.\ 1995; Weaver,
Krolik \& Pier 1998; Matt 1999; Perola et al.\ 1999), with the obtained
e-folding energies being in fair agreement with those obtained for the average
spectra. The e-folded power law, however, provides only a rough approximation
to the actual spectral shape from thermal Comptonization, and the relationship
between the e-folding energy and the electron temperature is not
straightforward (e.g., Stern et al.\ 1995).

The purpose of this work is to determine the tightest currently possible
constraints on the range of electron temperature, $kT$, and the Thomson optical
depth, $\tau$, in radio-quiet Seyferts. To achieve this, we compute here the
average OSSE spectra of {\it all\/} radio-quiet Seyfert 1s and of Seyfert 2s
observed (except NGC 4151, the brightest object by far). We stress that OSSE
still remains the most sensitive detector in the range $\ga 100$ keV currently
operating and it will be surpassed only by the IBIS detector aboard {\it
INTEGRAL\/} (Ubertini et al.\ 1999).

We fit the obtained data sets (with the useful energy range of $\sim 50$--500
keV) with a highly-accurate model of thermal Comptonization of Poutanen \&
Svensson (1996). We find that their statistical accuracy is sufficient to
provide useful constraints on the average parameters of the Comptonizing plasma
in those sources. Furthermore, we apply this model to the previously obtained
average OSSE spectrum of NGC 4151 (J97) and to the average $\sim 2$--500 keV
spectra of Seyfert 1s from \ginga\/ and OSSE and \exosat\/ and OSSE of G96.

\section{The data}
\label{s:data}

Table 1 gives the log of observations of all (except NGC 4151) radio-quiet
Seyfert 1s and 2s detected by OSSE through the end of 1998. The data for
individual objects include systematic errors estimated from the uncertainties
in the low-energy calibration and response of the detectors using both in-orbit
and pre-launch calibration data, and correspond to an uncertainty in the
effective area in the OSSE response. This uncertainty is $\sim 3\%$ at 50 keV
decreasing to $\sim 0.3\%$ at $\ga 150$ keV.

\begin{figure} \plotone{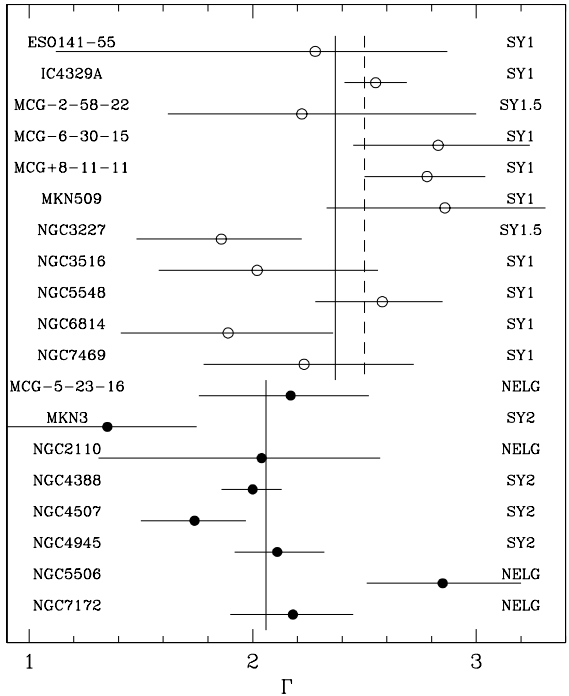}
\caption{\small  The distribution of 50--200 keV spectral indices in Seyfert 1s
(open circles) and Seyfert 2s (filled circles). The upper vertical solid and dashed lines show the mean and weighted mean for Seyfert 1s, respectively, and the lower solid line corresponds to both means for Seyfert 2s. The error bars are $1\sigma$. The general classes of Seyfert 1 and 2 include those classified as Seyfert 1.5 and NELG (narrow emission-line galaxy), respectively, according to the NASA Extragalactic Database.
}
\label{indices} \end{figure}

Table 1 also gives the photon power-law indices, $\Gamma$, fitted in the 
50--200 keV range to sums of observations of individual Seyferts with detection 
significance $\ga 4\sigma$. The indices are plotted in Figure \ref{indices}. 
The mean and its error of the best-fit values for Seyfert 1s and 2s are 
$2.37\pm 0.11$ and $2.06\pm 0.15$, respectively. The intrinsic dispersions of 
the distributions (given by their standard deviations) are 0.37 and 0.42, 
respectively. The probability that the 2 distributions have the same mean and 
standard deviation is 5\%, as obtained from the Student $t$ distribution. The 
mean weighted by the uncertainties and its error for Seyfert 1s and 2s are 
$2.50\pm 0.09$ and $2.05\pm 0.09$, respectively. 

We have then coadded  the spectra of Seyfert 1s and 2s, obtaining data sets
with $1.3\times 10^7$ s and $9.1\times 10^6$ s of OSSE exposure (scaled to a
single OSSE detector with an effective area of 500 cm$^2$) with $2.8\times
10^6$ and $2.7\times 10^8$ of source photons (in the 50--150 keV range) from 17
and 10 AGNs, respectively. Source photons in the average Seyfert 1 and 2
spectra are then detected in the ranges of $\sim 50$--500 keV and 50--400 keV,
respectively. The upper limits at higher energies are much above our model
spectra and thus we use only the above energy ranges in our fits.

\section{Spectral fits to the average spectra}

First, we note that the average OSSE data when fitted by a power law (PL) give
spectral indices, $\Gamma$, significantly higher (Table 2) than the average
X-ray spectral index of Seyfert 1s, $\Gamma_{\rm X}\simeq 1.9$--2.0 (Nandra \&
Pounds 1994). This implies the presence of a spectral break between $\sim 20$
and $\sim 50$ keV. Then, we test for the presence of  high-energy cutoffs in
the average spectra themselves.  We find statistically significant fit
improvements when the power-law model is replaced by an e-folded power law
(EPL) in both Seyfert 1s and 2s, see Table 2. The probability that the fit
improvement were by chance (obtained using the F-test) equals to 0.003 in each
case. This result argues strongly that spectra of individual Seyferts have
high-energy cutoffs in a relatively narrow range of energies.  If, instead,
there were a wide power-law distribution of the cutoff energies, the sum
spectrum would be a power law without a cutoff.

\begin{figure} \plotone{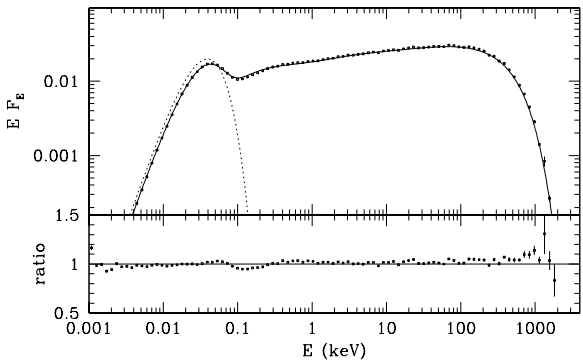} \caption{\small A comparison the
thermal-Comptonization spectrum from {\tt compps} (solid curve) with the
corresponding one obtained using a Monte Carlo method (points with error bars).
The lower panel shows the ratio between the latter and the former. The
Comptonizing plasma forms a sphere with the electron temperature of $kT=178$
keV and the radial optical depth of $\tau=0.43$ with the source of blackbody
seed photons at $kT_{\rm bb}=10$ eV (dotted curve) at its center. }
\label{compare} \end{figure}

We then consider a model with thermal Comptonization (TC) and Compton
reflection. We use the model {\tt compps} v3.4\footnote{{\tt compps} code is
available on the internet at ftp://ftp.astro.su.se/pub/juri/XSPEC/COMPPS}
(Poutanen \& Svensson 1996) in {\sc xspec} (Arnaud 1996). Since the actual
geometry of Seyferts remains mostly unknown, we choose here spherical geometry,
which is relatively generic, with spectra independent of the viewing angle. In
similar spirit, we assume the source of seed photons at the center with a
blackbody spectrum at $kT_{\rm bb}=10$ eV. This model has been extensively
tested against a Monte Carlo method (Gierli\'nski 2000), and Figure
\ref{compare} shows a good agreement between spectra obtained with the two
methods (for the plasma parameters close to those of the \ginga-OSSE fit
below). As the independent parameters of the model, we choose the electron
temperature, $kT$, and the Compton $y$ parameter,
\begin{equation} y\equiv
4\tau {kT\over m_{\rm e} c^2}, \end{equation}
where $\tau$ is the radial Thomson optical depth. A given value of $y$
corresponds to an approximately constant value of the 2--10 keV spectral index,
$\Gamma_{\rm X}$ (e.g., Ghisellini \& Haardt 1994; Poutanen 1998; Beloborodov
1999b; see below).

Reflection is treated using angle-dependent Green's functions of Magdziarz \&
Zdziarski (1995) and neglecting Comptonization of the reflected radiation in
the hot plasma. Then, the obtained strength of reflection, $R$ (defined
relative to the reflection strength from an isotropic point source above a
slab), corresponds approximately to its unscattered fraction. Since the OSSE
data cannot independently constrain $y$, $kT$ and $R$, we keep $R$ fixed at
0.75, obtained by fitting the average Seyfert-1 spectrum from \ginga\/ and OSSE
of G96. The inclination assumed by G96 is $\cos i=0.87$ (e.g.\ Nandra et al.\
1997), which we also assume in our fit to the Seyfert-1 spectrum. On the other
hand, Seyfert 2s are most likely seen more edge-on, and we assume $\cos i=0.4$
for the Seyfert-2 spectrum. We assume the reflecting medium is close to neutral
with the abundances of Anders \& Ebihara (1982).

\begin{figure} \plotone{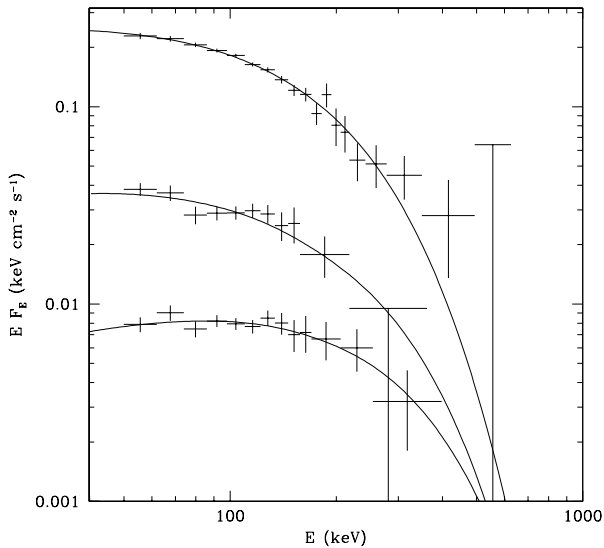} \caption{\small The average OSSE spectra
of NGC 4151, Seyfert 1s and Seyfert 2s (rescaled by a factor $1/5$), from top
to bottom, fitted by thermal Comptonization in a spherical cloud and Compton
reflection (with the assumed $R=0.4$ and 0.75 for NGC 4151 and Seyferts,
respectively).  } \label{spectra} \end{figure}

\begin{figure} \plotone{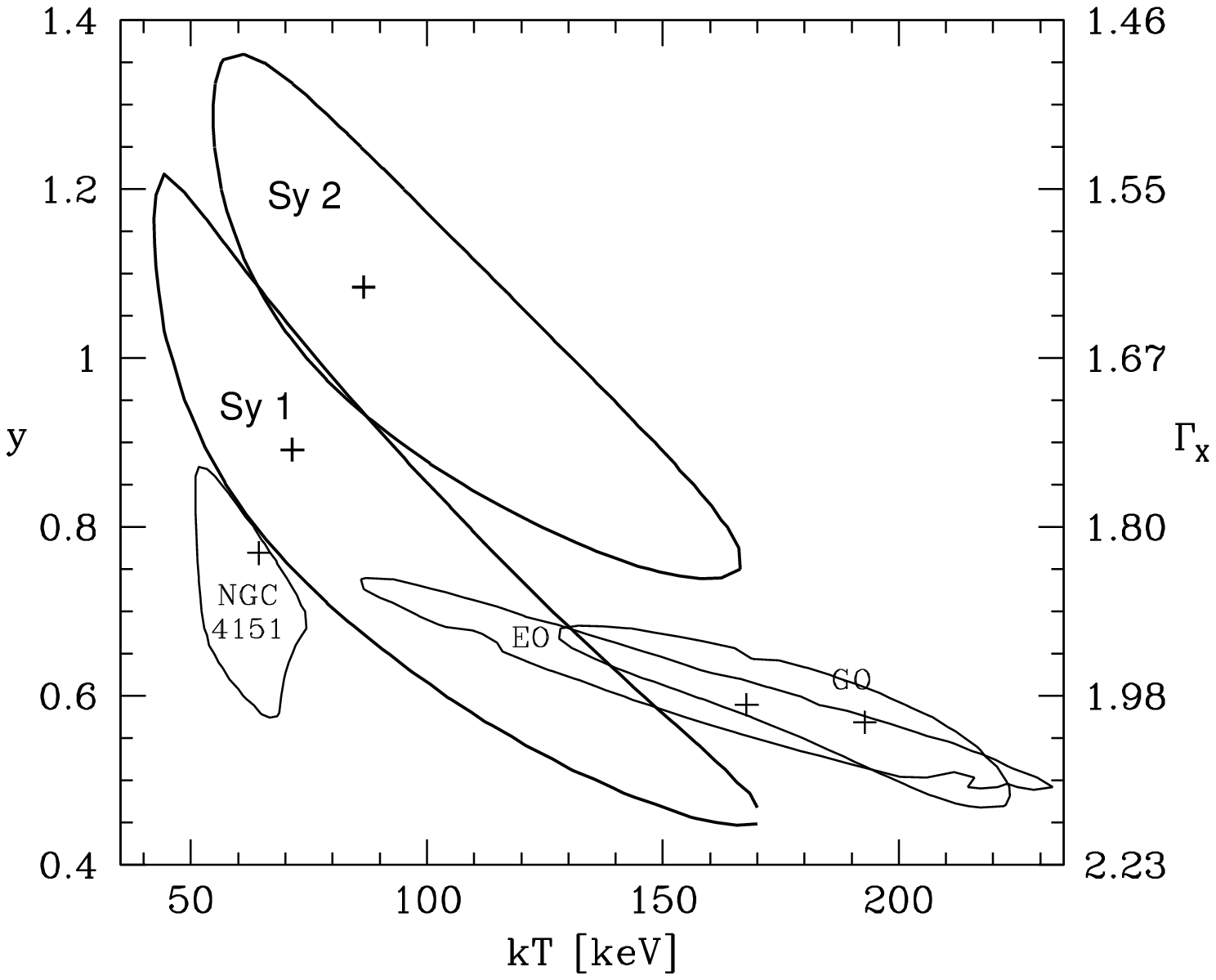} \caption{\small The $1\sigma$ error
contours for the average OSSE spectra of Seyfert 1s and 2s fitted by thermal
Comptonization in spherical geometry and Compton reflection.  The contours for
NGC 4151 and the average \ginga-OSSE (marked 'GO') and \exosat-OSSE (marked
'EO') spectra of G96 are also shown. The right vertical axes in this figure and
Figure \ref{slab} give the 2--10 keV intrinsic spectral index for $kT=100$
keV.}
\label{contours} \end{figure}

The resulting data and model spectra are shown in Figure \ref{spectra}. The
parameters are given in Table 2 and the error contours are shown in Figure
\ref{contours}. Its right vertical axis shows the value of the index
$\Gamma_{\rm X}$ (of the Comptonization spectrum without the reflection
component) computed at $kT=100$ keV. As mentioned above, $\Gamma_{\rm X}$ is a
weak function of $kT$ at a given $y$, and it changes within $\Delta\Gamma_{\rm
X}\la \pm 0.1$ for 50 keV $\la kT\la 200$ keV.

We see that the error contour for Seyfert 1s is consistent within $1\sigma$
with their average X-ray spectral index being approximately $\Gamma_{\rm
X}\simeq 1.9\pm 0.1$ (Nandra \& Pounds 1994; Zdziarski et al.\ 1999). The range
of $kT$ corresponding to this $\Gamma_{\rm X}$ is $\sim 60$--150 keV.

We have also refitted the average X-ray/\g-ray (X$\gamma$) spectra of Seyfert 
1s observed by both {\it Ginga\/} and OSSE and by \exosat\/ and OSSE of G96. We 
fitted them (in the range $\leq 500$ keV) in the same way as in G96 except for 
modeling the primary continuum by thermal Comptonization instead of the e-
folded power law. We obtain $kT=190$ keV, 166 keV, and $y=0.57$, 0.59, 
respectively, at the best fits, with the error contours shown in Figure 
\ref{contours}. The resulting parameters are consistent within $1\sigma$ with 
those of the average spectrum of all Seyfert 1s observed by OSSE. However, our 
current data set, with photon statistics being much better than those of the 
two OSSE spectra of G96, represents the currently best estimate of the average 
$kT$ in Seyfert 1s.

Figures \ref{spectra} and \ref{contours} also show the spectrum and the error
contour, respectively, for the average of OSSE observations of the brightest
Seyfert, NGC 4151 (J97) obtained with the same model. Since
the strength of reflection in NGC 4151 is rather uncertain due to strong
absorption in X-rays, we have constrained $R$ to be in the 0--1 range while
computing the error contour (which assumption is responsible for its irregular
shape). We see that the X-ray spectrum implied by these data is very similar to
that of typical Seyfert 1s, with $\Gamma_{\rm X}\sim 1.8$--2.0. The obtained
range of temperature is relatively low, $kT\sim 50$--80 keV,  in agreement with
the results of Zdziarski et al.\ (1996) and J97.

\section{Comparison between Seyfert 1s and 2s}

The weighted average of the 50--200 keV indices of individual objects (Table 1 
and Figure \ref{indices}) and the indices in the power-law fits to the sum 
spectra (Table 2) for Seyfert 1s, $2.50\pm 0.09$ and $2.56\pm 0.14$, 
respectively, are significantly softer than those for Seyfert 2s, $2.05\pm 
0.09$ and $2.21\pm 0.12$. The probability that the 2 samples are drawn from the 
same distribution is only 5\% (\S \ref{s:data}). This difference is confirmed 
by Figures \ref{spectra}, \ref{contours}, where we see that the average 
spectrum of Seyfert 2s is noticeably harder than that of Seyfert 1s.

We have investigated whether this difference can be explained by the viewing
angle different between Seyfert 1s and 2s. One relevant effect is the strength
of Compton reflection decreasing with the increasing viewing angle. Since the
spectrum from Compton reflection typically peaks around 30 keV followed by a
steep decline at higher energies, the larger $R$ the softer the spectrum in the
OSSE range. However, this effect is already included in our
Comptonization/reflection model, and Figure \ref{contours} shows that it is not
sufficient to account for the difference between the $y$ parameters of the
average spectra. We note, however, that the average inclination of Seyfert 2s
remains unknown. We have found we can fit the two spectra with the same
Comptonization/reflection model (with $R=0.75$) if $\cos i=0^{+0.3}$ for
Seyfert 2s within 90\% confidence.

\begin{figure}
\plotone{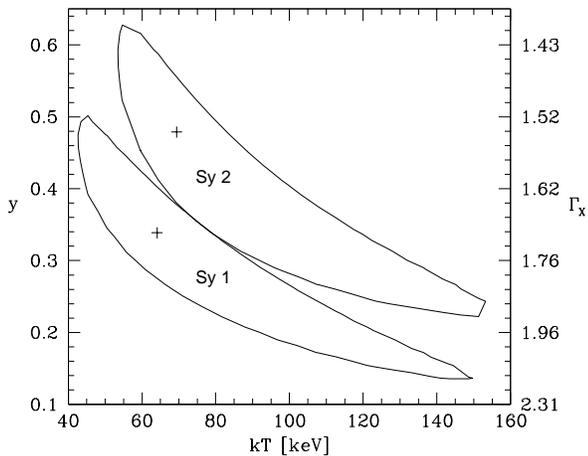} \caption{\small The $1\sigma$ error contours for the
average OSSE spectra of Seyfert 1s and 2s in the case of thermal Comptonization
in a slab.  } \label{slab} \end{figure}

An additional subtle effect appears when the Comptonizing medium has a slab
geometry. Namely, photons emitted at a large viewing angle (with respect to the
slab normal) undergo a larger number of scatterings than those emitted at a
small viewing angle, due to the escape probability at a given depth from the
surface, $\tau'$, in a given direction being $\exp(-\tau'/\cos i)$. We have
thus fitted our data with a model in which the Comptonizing medium forms a slab
with a half-thickness $\tau$ and the seed photons are emitted by point sources
in the midplane. We have first fitted with this model the \ginga-OSSE spectrum
of G96, and then fixed the obtained $R$ in the fits to our average spectra. The
resulting best-fit parameters are $kT=63$ keV, 67 keV, $y=0.34$, 0.48, for
Seyfert 1s and 2s, respectively. The error contours are shown in Figure
\ref{slab}. We see that the two parameter ranges are still incompatible with
each other. This means that the spectral shape does not change sufficiently
due to this effect to account for the observed spectral difference (see
Figure \ref{spectra}). For example, in the case of the best-fit Seyfert-1 model
spectrum (including both Comptonization and reflection), the ratio of
monochromatic fluxes at 150 keV and 50 keV increases only by 14\% when the
inclination changes from $\cos i=0.87$ to $\cos i=0.4$.

The above results also show that the obtained values of $kT$ and $\Gamma_{\rm
X}$ only weakly depend on geometry. For Seyfert 2s, our results imply
$\Gamma_{\rm X}$ between 1.4 and 1.9 and $kT$ between 50 and 170 keV.

We then consider possible effects of absorption/scattering in a torus 
surrounding the central source being important in Seyfert 2s. This can be an 
important effect in the OSSE range only when the torus is Thomson-thick, 
$N_{\rm H} \ga 1.5\times 10^{24}$ cm$^{-2}$. Then the hardening occurs (apart 
from a weak effect of photo-electric absorption) because the torus becomes more 
transparent to scattering with the increasing photon energy due to the 
Klein-Nishina effects. We have applied a numerical model of torus 
absorption/scattering of Krolik, Madau, \& \.Zycki (1994) to our average 
spectrum of Seyfert 2s assuming the intrinsic spectrum from thermal 
Comptonization of Seyfert 1s. We have obtained a good fit ($\chi_\nu^2=22/26$) 
at $N_{\rm H}=3\times 10^{24}$ cm$^{-2}$ with an intrinsic TC spectrum 
identical in the cases of Seyfert 1s and 2s with a torus with an opening angle 
of $60\degr$. However, since that numerical model does not allow for including 
Compton reflection from a disk, we cannot obtain meaningful constraints on its 
parameter space.

On the other hand, the only Seyfert 2 in our sample with a Thomson-thick
absorber is NGC 4945 (Risaliti, Maiolino \& Salvati 1999), with $N_{\rm
H}\simeq 4\times 10^{24}$ cm$^{-2}$ (see also Madejski et al.\ 2000), which
nucleus is also viewed edge-on (Greenhill, Moran \& Herrnstein 1997). Also,
some effect of absorption on the OSSE band is possible in Mkn 3, with
$N_{\rm H}\simeq 1.3\times 10^{24}$ cm$^{-2}$ (Cappi et al.\ 1999). Then,
$N_{\rm H}\sim (3$--$4)\times 10^{23}$ cm$^{-2}$ in NGC 4507 and NGC 4388, and
$< 10^{23}$ cm$^{- 2}$ in other Seyfert 2s in our sample (Risaliti et al.\
1999), in which cases the effect of scattering on our spectra is negligible.

Thus, we cannot rule out an intrinsic difference between Seyfert 1s and 2s. 
Such a difference may be hinted for by X-ray spectra of individual Seyfert 2s 
often appearing harder than those of Seyfert 1s (e.g., Bassani et al.\ 1995; 
Smith \& Done 1996). If indeed the X-ray spectra of Seyfert 2s were harder than 
those of Seyfert 1s, then our results above would indicate a correlation 
between the X-ray slope and that in the OSSE band. On the other hand, we 
consider it possible that a complex structure of the absorber causes 
underestimating of the actual $\Gamma_{\rm X}$ and $N_{\rm H}$ in some Seyfert 2s. 

Finally, we consider the possibility that our obtained difference between the
soft \g-ray spectra of Seyfert 1s and 2s is simply caused by large relative
contributions of a few bright objects to our average spectra. In the case of
Seyfert 1s, 38\% of the total number of source photons come from IC 4329A. For
Seyfert 2s, we have 29\% and 24\% contributions from NGC 4388 and NGC 4945,
respectively. Thus, if those objects had atypical spectra for their categories,
they would significantly bias the shape of the average spectra. However,
inspection of the distribution of the spectral indices in Figure \ref{indices}
does not confirm this supposition. All 3 objects have spectra quite typical for
their corresponding category.

On the other hand, it is interesting that 4 narrow-emission line galaxies
(NELGs) in our sample of Seyfert 2s (Fig.\ \ref{indices}) have somewhat softer
spectra than regular Seyfert 2s (which effect is most pronounced for NGC 5506).
NELGs are objects intermediate between Seyfert 1s and 2s, and are often
considered together with Seyfert 1s in one class (e.g., Nandra \& Pounds 1994).
Figure \ref{indices} also marks 2 Seyferts classified as type 1.5 included in
our Seyfert-1 sample.

In order to further investigate the issue of the effect of individual objects 
on our average spectra, we have also studied average Seyfert 1 and 2 spectra 
derived from observations up to the end of 1995. Those averages were based on 
$\sim 2/3$ of the statistics of the present spectra ($1.8\times 10^6$ and 
$1.6\times 10^6$ photons, respectively). Those samples contain, e.g., less than 
half of the photons from IC 4329A, none of NGC 3516, and almost none from NGC 
4945 (see Table 1). Thus, if IC 4329A and NGC 4945 were substantially different 
from the average Seyfert 1 and 2, respectively, we would expect significant 
differences in the spectral shape of the old and new averages. However, the old 
and new average spectra can be fitted with identical models with hardly any 
increase of $\chi^2$ with respect to independent fits, and the F-test gives the 
probability that their spectral shape differ of $\la 50\%$ for either Seyfert 
1s or 2s.

Summarizing, we do find that Seyfert 2s have significantly harder spectra than 
Seyfert 1s (whereas their cutoff energies or temperatures are similar). This 
difference can be explained by the dependence of Compton reflection on 
orientation only for orientation close to edge-on ($\cos i\la 0.3$). The 
angular dependence of thermal-Comptonization spectra in slab geometry is of 
relatively minor importance. The effect of Thomson-thick absorption in a torus 
surrounding the X$\gamma$ source is important in NGC 4945 and, possibly, 
Mkn 3.

\section{Discussion}

We have obtained quantitative constraints on the average electron temperature
and the Compton parameter (or, equivalently, the Thomson optical depth) in
Seyfert 1s and 2s (approximately $kT\sim 50$--150 keV and $y<1$, $\tau\la 1$).
These results put constraints on theoretical models of Seyferts as well as of
the cosmic X$\gamma$ background. A detailed analysis of these constraints is
beyond the scope of this work and we only outline the main relevant issues
below.

Physically, the value of temperature in a source is determined by energy
balance between heating and cooling. The balance depends, first of all, on the
source geometry, which determines the flux of seed photons incident on the
plasma. Production of seed photons appears to be dominated by blackbody photons
emitted by an optically-thick medium in the vicinity of the hot plasma
(Zdziarski et al.\ 1999), with the flux of thermal synchrotron photons being
negligible in luminous AGNs (Wardzi\'nski \& Zdziarski 2000). Thus, the values
of $kT$ puts constraints on the geometry of the X-ray sources. Two geometries
appear possible: a patchy corona above a cold accretion disk, and a hot
accretion disk with an overlapping cold medium (either an outer cold disk, cold
blobs or both, see, e.g., Poutanen 1998, Zdziarski et al. 1998).

The patchy corona geometry has been discussed recently by, e.g.,  Haardt,
Maraschi \& Ghisellini (1994), Stern et al.\ (1995), Poutanen \& Svensson
(1996), G96, and Beloborodov (1999a). In the case of Seyferts which have on
average quite soft spectra ($\Gamma_{\rm X}\sim 1.9$) and large reflection
($R\sim 0.75$), the patchy static corona still appears to be a viable model.
The situation is different for those objects which have hard spectra and little
reflection (see Zdziarski et al.\ 1999). In order to produce hard spectra, the
emitting region  should be well separated from the cold accretion disk (e.g.,
Svensson 1996), but in that case the predicted reflection is close to unity.
Mildly relativistic motions of emitting plasma away from the disk, however,
solve both problems producing hard spectra and little reflection (Beloborodov
1999a, b). We also note that this model can produce reflection larger than
unity when emitting plasma is moving towards the disk.

Coronal models either with or without \ee\ pair production are possible. We
note that studies of thermal pair plasmas in pair equilibrium predict no
distinct pair annihilation even from a pair-dominated plasmas
(Macio{\l}ek-Nied\'zwiecki, Zdziarski \& Coppi 1995). Thus, the lack of such a
feature observed by OSSE (e.g., J97) does not rule out the presence of thermal
\ee\ pairs.

The hot disk model has been developed by Shapiro, Lightman \& Eardley (1976),
whose solution branch was cooling-dominated (and thermally unstable). Including
advection gives rise to a stable, low-luminosity solution branch,  and the
intersection of the two branches limits the luminosity and the optical depth of
an inner flow (Narayan \& Yi 1995; Abramowicz et al.\ 1995; see Zdziarski 1998
for a model parametrizing the flow by $y$). The role of \ee\ pair production is
in general negligible (Bj\"ornsson et al.\ 1996). The values of $kT\sim 100$
keV and $\tau\la 1$ are predicted by this model close to the maximum possible
luminosity of the hot flow, in an agreement with our results. In order to
account for the observed range of $\Gamma_{\rm X}$ and $R$, an overlap between
the hot disk and a cold one is required (Poutanen, Krolik \& Ryde 1997;
Zdziarski et al.\ 1999).

It is noteworthy that black-hole binaries in their hard states show X$\gamma$
spectra very similar to those of radio-quiet Seyferts. For example, spectra of
Cyg X-1 and GX 339--4 have been well modeled by thermal Comptonization at
$\tau\sim 1$, and $kT\sim 100$ keV and $\sim 60$ keV, respectively
(Gierli\'nski et al.\ 1997; Zdziarski et al.\ 1998). Hard-state X$\gamma$
spectra of X-ray novae are also similar (Grove et al.\ 1998).

Another class of objects for which a comparison with Seyferts is of interest is 
broad-line radio galaxies. Their X$\gamma$ spectra have been studied by 
Wo\'zniak et al.\ (1998), who found both their X-ray and soft \g-ray spectra 
(with $\langle\Gamma_{\rm X}\rangle = 1.67\pm 0.18$, and  $\langle\Gamma\rangle 
= 2.15\pm 0.16$ in the 50--500 keV band) to be harder than those of Seyfert 1s 
(similarly to the case of Seyfert 2s). On the other hand, Wo\'zniak et al.\ 
(1998) have not found a high-energy cutoff in their average OSSE spectrum, 
indicating the importance of non-thermal Compton scattering. This conclusion 
has to be considered tentative as the statistical accuracy of that spectrum is 
much below those of radio-quiet Seyferts. Still, a support for the importance 
of non-thermal processes in radio galaxies is provided by the \g-ray spectrum 
of the radio galaxy Cen A, which has a broken power-law form with the spectral 
breaks at $\sim 100$ keV and $\sim 10$ MeV (Steinle et al.\ 1998).

The form of the average spectra of Seyfert 1s and 2s is of primary importance
synthesizing the cosmic X$\gamma$ background from individual sources. This
synthesis appears possible provided Seyfert 2s dominate the hard X-ray
background (e.g., Zdziarski et al.\ 1995; Comastri et al.\ 1995).

\section{Conclusions}

We have obtained error contours in the $kT$-$y$ space for the average spectra
of Seyfert 1s and 2s observed by OSSE. Combining the contours for Seyfert 1s
with their average X-ray spectral index of $\Gamma_{\rm X}\simeq 1.9$--2
(Nandra \& Pounds 1994), we obtain $kT\simeq 50$--150 keV. The corresponding
Thomson optical depth ($\tau\simeq y 128\,{\rm keV}/kT$) of Seyfert 1s is
$\tau\simeq 0.3$--1.5 in spherical geometry and $\tau\simeq 0.1$--0.6 in slab
geometry.

We also find both the average and individual Seyfert-2 spectra are 
significantly harder than those of Seyfert 1s. This difference can be partly 
due to the spectral component from Compton reflection being weak in Seyfert 2s, 
provided their average inclination is edge-on enough. Furthermore, 
Thomson-thick absorption can account for the difference in 2 (out of 8) objects in our sample.

\acknowledgements

This research has been supported in part by a grant from the Foundation for 
Polish Science and KBN grants 2P03C00511p0(1,4) and 2P03D00614, NASA grants and 
contracts, the Swedish Natural Science Research Council, and the Anna-Greta and 
Holger Crafoord Fund. We thank Marek Gierli\'nski for his assistance with 
implementing models into the {\sc xspec} software package and with Monte Carlo 
calculations, Piotr \.Zycki for providing his Thomson-thick torus model, and 
the referee for insightful remarks.

\begin{deluxetable}{lccc}
\tablecolumns{4}
\tablewidth{0pc}
\tablecaption{OSSE observations of Seyferts\tablenotemark{a}}
\tablehead{
\colhead{Object} & \colhead{$\Gamma$\tablenotemark{b}\, or Dates}   &
\colhead{Exposure\tablenotemark{c}}
& \colhead{Photon flux\tablenotemark{d}} }
\startdata
\cutinhead{Seyfert 1s}
ESO 141--55 & $2.28^{+0.59}_{-1.16}$  &3.10  &$2.63\pm 0.79$\\
                      &92/220--92/223&0.90  &$3.57\pm 1.42$\\
                      &92/241--92/245&1.03  &$3.07\pm 1.35$\\
                      &92/290--92/303&1.17  &$1.53\pm 1.31$\\
IC 4329A       &  $2.55^{+0.14}_{-0.14}$ &27.44&$7.74\pm 0.40$\\
                      &92/283--92/289&1.10  &$7.35\pm 1.64$\\
                      &92/309--92/322&2.77  &$7.93\pm 1.01$\\
                      &93/013--93/033&4.17  &$8.93\pm 0.66$\\
                      &94/305--94/313&1.08  &$3.64\pm 2.78$\\
                      &95/024--95/045&4.13  &$6.93\pm 0.96$\\
                      &96/212--96/226&2.52  &$6.85\pm 1.24$\\
                      &96/226--96/233&1.40  &$6.94\pm 1.65$\\
                      &96/233--96/240&1.67  &$7.52\pm 1.52$\\
                      &96/303--96/317&2.93  &$11.61\pm 1.00$\\
                      &97/217--97/230&5.68  &$6.85\pm 1.20$\\
III Zw II          &     ---             &4.17 &$1.63\pm 0.79$\\
                      &93/083--93/088&0.61  &$<3.21$\\
                      &96/065--96/078&3.57  &$1.97\pm 0.88$\\      
MCG --2-58-22&  $2.22^{+0.78}_{-0.60}$     &2.95  &$3.72\pm 0.88$\\
                      &93/053--93/056&0.59  &$<3.40$\\
                      &93/088--93/091&0.43  &$2.86\pm 2.11$\\
                      &93/125--93/126&0.38  &$3.94\pm 2.11$\\
                      &95/094--95/101&1.55  &$5.27\pm 1.32$\\
MCG --6-30-15  &  $2.83^{+0.41}_{-0.38}$ &5.07  &$4.58\pm 0.67$\\
                      &92/283--92/289&1.11  &$3.91\pm 1.15$\\
                      &92/309--92/322&2.75  &$4.80\pm 0.71$\\
                      &95/164--95/171&1.21  &$4.68\pm 2.04$\\
MCG +8-11-11  & $2.78^{+0.26}_{-0.28}$ &14.61 &$4.13 \pm 0.39$\\
                      &92/163--92/177&6.70   &$4.41 \pm 0.51$\\   
                      &93/145--93/151&1.94   &$2.50 \pm 0.94$\\   
                      &95/234--95/250&5.97   &$4.33 \pm 0.68$\\
Mkn 279       &          ---          &5.70    &$1.57 \pm 0.58$\\
                      &92/066--92/079&4.14    &$  2.47 \pm 0.66$\\   
                      &93/230--93/236&1.56    &$   < 2.35      $\\
Mkn 509       &  $2.86^{+0.45}_{-0.53}$  &4.88&$       3.98 \pm 0.70$\\
                      &92/304--92/308&1.04&$      4.22 \pm 1.12$\\
                      &93/082--93/088&0.97&$     3.30 \pm 1.37$\\
                      &94/138--94/144 &0.10&$      <21.76      $\\
                      &95/010--95/024 &2.77&$     3.98 \pm 0.98$\\
Mkn 841       &    ---    &2.64&$      3.56 \pm 1.02$\\
                      &92/108--92/114 &0.26&$      < 5.15      $\\
                      &96/277--96/289  &2.38&$    3.79 \pm 1.09$\\
NGC 3227      &  $1.86^{+0.36}_{-0.38}$ & 10.97&$        3.40 \pm 0.55$\\
                      &95/003--95/010&0.40&$     4.09 \pm 3.03$\\
                      &95/220--95/234&2.42&$      3.09 \pm 1.49$\\
                      &95/234--95/250&5.90&$     2.84 \pm 0.67$\\
                      &95/290--95/304&2.25&$      5.08 \pm 1.10$\\
NGC 3516      &   $2.02^{+0.56}_{-0.44}$ &10.50&$        2.97 \pm 0.64$\\
                     &97/175--97/195&5.07&$      2.38 \pm 0.88$\\
                     &97/245--97/252&1.64&$      5.15 \pm 1.67$\\
                     &97/266--97/280&3.12&$     2.67 \pm 1.18$\\
                     &98/013--98/021&0.67&$      3.45 \pm 2.92$\\
NGC 3783      & ---                   &0.98&$     4.42 \pm 1.38$\\
                     &92/178--92/184 &0.98&$     4.42 \pm 1.38$\\
NGC 526A      &                  ---          &3.44&$    3.69 \pm 0.91$\\
                        &   96/065--96/078  &3.44&$    3.69 \pm 0.91$\\
NGC 5548      &   $2.58^{+0.27}_{-0.30}$  &11.15&$  5.10 \pm 0.54$\\
                       &91/228--91/234&1.82&$      4.24 \pm 1.22$\\ 
                      &91/291--91/303&0.57&$      5.04 \pm 1.94$\\   
                      &91/305--91/310&0.67&$      6.73 \pm 2.14$\\    
                     &93/250--93/252&0.53&$      5.88 \pm 1.89$\\
                     &93/264--93/265&0.11&$      4.97 \pm 4.03$\\
                     &93/265--93/274&1.58&$      4.42 \pm 1.13$\\
                     &93/291--93/292&0.55&$      8.85 \pm 1.98$\\
                     &95/270--95/276&1.07&$      4.56 \pm 1.36$\\
                     &95/311--95/318&1.58&$      5.51 \pm 2.07$\\
                     &96/081--96/094&2.68&$      4.74 \pm 1.13$\\
NGC 6814      & $1.89^{+0.47}_{-0.48}$ &5.05&$             3.01 \pm 0.58$\\
                     &93/033--93/040&2.54&$      3.15 \pm 0.82$\\
                     &93/215--93/222&2.52&$      2.88 \pm 0.83$\\ 
NGC 7213      & --- &5.15&$         2.12 \pm 0.86$\\
                     &93/328--93/335&0.90&$       < 3.81      $\\
                     &94/032--94/039&0.47&$      5.13 \pm 2.99$\\
                     &95/010--95/024&2.14&$      3.52 \pm 1.15$\\
                     &96/205--96/212&1.64&$      3.35 \pm 1.78$\\
NGC 7469      & $2.23^{+0.49}_{-0.45}$  &9.24&$          3.43 \pm 0.62$\\
                     &94/067--94/074&2.60&$      3.63 \pm 0.94$\\
                     &94/109--94/116&2.89&$      3.92 \pm 0.89$\\
                     &94/137--94/144&2.01&$      4.63 \pm 1.76$\\
                     &94/213--94/216&0.91&$       < 4.78      $\\
                     &96/081--96/094&0.84&$      3.20 \pm 1.94$\\
\cutinhead{Seyfert 2s}
MCG --5-23-16&  $2.17^{+0.35}_{-0.41}$ &7.72&$         4.26 \pm 0.55  $\\
                      &92/219--92/224&1.77&$      4.53 \pm 1.02  $\\ 
                      &92/241--92/245 &1.07&$     4.32 \pm 1.33  $\\
                     &94/216--94/221&0.56&$      4.12 \pm 2.41  $\\
                     &94/263--94/277&4.32&$      4.15 \pm 0.77  $\\
Mkn 3 & $1.35^{+0.40}_{-0.94}$ &9.65&$    1.87 \pm 0.51  $\\
                     &94/074--94/081&3.12&$      1.59 \pm 0.87  $\\
                     &94/213--94/216&1.06&$       < 3.02        $\\
                     &95/045--95/052 &2.62&$     2.71 \pm 0.98  $\\
                     &95/052--95/059 &2.85&$     3.19 \pm 0.94  $\\
NGC 1275 & --- &6.57&$                                  1.97 \pm 0.52  $\\
                      &91/333--91/346&4.46&$      1.58 \pm 0.55  $\\
                     &94/116--94/130&2.11&$      2.79 \pm 1.13  $\\
NGC 2110         &$2.04^{+0.43}_{-0.73}$ &4.48&$      4.06 \pm 0.81  $\\
         &96/149--96/163&4.48&$      4.06 \pm 0.81  $\\
NGC 4388& $2.00^{+0.13}_{-0.14}$ &17.75&$                 8.82 \pm 0.45  $\\
                      &92/261--92/282  &5.31&$    8.13 \pm 0.74  $\\
                     &93/238--93/250&1.67&$     13.78 \pm 1.12  $\\
                     &95/220--95/234&2.91&$      7.98 \pm 1.06  $\\
                     &95/290--95/304 &2.75&$    13.53 \pm 1.11  $\\
                     &97/203--97/217      &2.71&$5.85 \pm 1.39  $\\
                     &97/280--97/293 &2.40&$     5.86 \pm 1.36  $\\
NGC 4507   & $1.74^{+0.23}_{-0.24}$ &8.88&$ 4.66 \pm 0.58  $\\
                     &93/033--93/040 &1.54&$     6.32 \pm 1.08  $\\
                     &93/215--93/222  &1.36&$    6.57 \pm 1.16  $\\
                     &96/289--96/303  &2.62&$    3.84 \pm 1.26  $\\
                     &97/105--97/126&3.36&$      3.77 \pm 0.95  $\\
NGC 4945              &$2.11^{+0.21}_{-0.19}$ &18.50&$  6.86 \pm 0.49  $\\
                     &94/305--94/313&1.31&$     11.52 \pm 1.19  $\\
                     &96/002--96/005    &1.10&$  8.54 \pm 1.59  $\\
                     &96/005--96/016&3.98&$     13.05 \pm 1.25  $\\
                     &97/280--97/307    &9.47&$  4.47 \pm 0.67  $\\
                     &98/314--98/320&2.64&$      3.09 \pm 1.24  $\\
NGC 5506& $2.85^{+0.35}_{-0.34}$ &7.00&$   5.72 \pm 0.72  $\\
                     &94/347--94/354 &2.18&$     5.06 \pm 1.10  $\\
                     &94/354--95/003 &3.21&$     5.88 \pm 0.91  $\\
                     &95/318--95/325&0.60&$       < 7.38        $\\
                     &95/332--95/341 &1.01&$     9.44 \pm 2.47  $\\
NGC 7172 &$2.18^{+0.27}_{-0.28}$ &8.84&$            7.10 \pm 0.72  $\\
                     &95/053--95/059&1.37&$      4.47 \pm 1.62  $\\
                     &95/059--95/066&1.42&$      4.71 \pm 1.35  $\\
                     &95/066--95/080&2.20&$      9.12 \pm 1.11  $\\
                     &96/038--96/044&2.68&$      8.42 \pm 1.37  $\\
                     &97/161--97/168&1.18&$      6.25 \pm 3.04  $\\
NGC 7582   & --- &4.48&$                                     3.77 \pm 0.98  $\\
                      &91/347--91/361&2.70&$      5.11 \pm 1.44  $\\
                      &92/093--92/100&0.59&$      4.83 \pm 1.61  $\\
                      &92/100--92/107&0.26&$       < 7.65        $\\
                      &92/107--92/114&0.30&$       < 4.50        $\\
                     &94/347--94/354&0.63&$       < 4.41        $\\
\enddata
\tablenotetext{a}{All uncertainties here are $1\sigma$; the first row for each
object corresponds to sum of all its observations.}
\tablenotetext{b}{The 50--200 keV photon index for objects with $\ga 4\sigma$
detection.}
\tablenotetext{c}{In units of $10^5$ s scaled to a single OSSE detector.}
\tablenotetext{d}{For 50--150 keV in units of $10^{-4}$ cm$^{-2}$ s$^{-1}$.}
\end{deluxetable}

\begin{deluxetable}{lccccc}
\tablecolumns{6}
\tablewidth{0pc}
\tablecaption{Spectral fits of the average OSSE spectra of Seyferts of type 1
and 2}
\tablehead{
\colhead{Model\tablenotemark{a}} & \colhead{Type}   &
\colhead{$F$\tablenotemark{b}}  & \colhead{$\Gamma$ or $y$\tablenotemark{c}}
&\colhead{$E_{\rm c}$ or $kT$\tablenotemark{d}} &\colhead{$\chi_\nu^2$} }
\startdata
PL  &1 &5.4 & $2.56^{+0.14}_{-0.14}$ &-- &30.9/35\\
    &2 & 7.0 &$2.21^{+0.12}_{-0.12}$ &-- &30.5/27\\
EPL &1 & 5.5 & $1.69^{+0.57}_{-0.81}$ & $120^{+220}_{-60}$  &23.8/34\\
       &2 & 7.0 & $1.33^{+0.56}_{-0.52}$ & $130^{+220}_{-50}$  &21.8/26\\
TC &1 & 5.4 &$0.89^{+0.36}_{-0.52}$ &$69^{+134}_{-28}$ &24.1/34\\
       &2 & 7.0 &$1.09^{+0.29}_{-0.41}$ &$84^{+101}_{-31}$  &21.7/26\\
\enddata
\tablenotetext{a}{The uncertainties correspond to 90\% confidence of 1
parameter.}
\tablenotetext{b}{Normalization given by the 50--150 keV flux in units of
$10^{-11}$ erg cm$^{-2}$ s$^{-1}$. }
\tablenotetext{c}{The photon index or the Compton parameter.}
\tablenotetext{d}{The e-folding energy or temperature in keV.}
\end{deluxetable}


\end{document}